\documentclass{article}
\usepackage{amssymb}
\usepackage{amsfonts}
\usepackage{aip}

\newtheorem{theorem}{Theorem}

\newtheorem{definition}[theorem]{Definition}

\input{tcilatex}

\begin{document}

\title{QS7: Perspectives: Quantum Mechanics on Phase Space}
\author{J. A. Brooke$^{\#}$ and F. E. Schroeck$^{\#\#}$ \\
$^{\#}$University of Saskatchewan, \\
Saskatoon, Canada;\\
brooke@sask.usask.ca\\
$^{\#\#}$University of Denver, \\
Denver, Colorado \\
and Florida Atlantic University,\\
Boca Raton, Florida, U.S.A.;\\
fschroec@du.edu\\
\textit{Dedicated to the memory of Eduard Prugove\v{c}ki (1937-2003)}}
\date{October 28, 2004}
\maketitle

\begin{abstract}
The basic ideas in the theory of Quantum Mechanics on Phase Space are
illustrated through an introduction of generalities which seem to underlie
most if not all such formulations and follow with examples taken primarily
from kinematical particle model descriptions exhibiting either Galileian or
Lorentzian symmetry. The structures of fundamental importance are the
relevant (Lie) groups of symmetries and their homogeneous (and associated)
spaces that, in the situations of interest, also possess Hamiltonian
structures. Comments are made on the relation between the theory outlined
and a recent paper by Carmeli, Cassinelli, Toigo, and Vacchini.


\end{abstract}

\section{Introduction}

The formulation of Quantum Mechanics on Phase Space, having origins as early
as the 1930s\cite{Weyl} and \cite{Wig32} , underwent something of a
resurgence in the late 1970s and early 1980s. A number of concepts, tools,
and elements introduced in the 1950s and 1960s in the theory of quantum
measurement (operator-valued measures with non-projector values being
perhaps the most significant), which today play an indispensible role in the
context of quantum computation and quantum information, have played an
equally critical role in theories of quantum mechanics on phase space. The
concepts of \textit{positive operator-valued measure (POVM)} and \textit{%
informational completeness} (of a collection of observables) are especially
worth mentioning in the phase space theories of quantum mechanics. 

The present paper is an amalgam of the talks by the two authors, and is
aimed at illustrating the basic ideas in the theory of Quantum Mechanics on
Phase Space through a "gentle" introduction of generalities which seem to
underlie most, if not all, such formulations.

As might be expected in any treatment of particle models embodying
non-relativistic or relativistic kinematics, the structures of fundamental
importance are the relevant (Lie) groups of symmetries and their respective
homogeneous (and associated) spaces which, in the situations of interest,
also possess Hamiltonian structures.

Quantum mechanical systems are \textit{characterizable} in terms of\ their
kinematical symmetries that, in their most basic form, are either Galileian
or Lorentzian, depending whether the system is required to incorporate
non-relativisic or relativisic principles. Since in quantum theory, the
description (and construction) of what may be called \textit{multi-particle
Hilbert spaces} derives (by suitable tensor products) from \textit{%
single-particle} formulations, one concentrates upon these elementary
systems. In Wigner's formulation in the relativistic context \cite{Wig39}
and L\'{e}vy-Leblond's non-relativistic counterpart \cite{Levy}, these are
associated with irreducible, unitary representations on the Hilbert space of
the system of the appropriate kinematic group: the Lorentz group or the
Galilei group, or the related inhomogeneous group (actually the
11-dimensional \textit{extended} version in the case of the Galilei group in
consequence of the fundamental work of Bargmann \cite{Barg}.) In both
situations, the \textit{physical} elementary systems are determined by two
real parameters, namely non-negative mass \textit{m} and spin \textit{j}
taking values that are non-negative half-integer multiples of Planck's
fundamental constant. It is now the \textit{physical interpretation} that
provides the guide through consideration of the \textit{classical mechanical}
phase spaces associated with the \textit{elementary, irreducible particle
models}. Since the pioneering works of Souriau\cite{Sour} and Kostant\cite%
{Kos}, it has been recognized that it is the symplectic homogeneous spaces
of the appropriate kinematical group that provide the correct phase space
descriptions, and that a particular \textit{model} of the phase space
picture (that embodies the covariance symmetry of the kinematical group)
arises either from the \textit{co-adjoint} representation of the group on
the dual of its Lie algebra or suitable extensions of the co-adjoint
representation approach to take account of topological aspects such as
non-trivial group cohomology as in the case of the inhomogeneous Galilei
group.

These assumptions are taken to be basic, and one proceeds to note the
consequences.

Note that the phase space formulation of quantum mechanics has little if
anything to do with the theory of \textit{geometric quantization} that
seeks, through the use of complex polarizations, to reduce the phase space
description to one involving a locally-Poisson-commuting collection of basic
coordinates. "Quantum Mechanics on Phase Space" is, in contrast, prepared to
accept the need to "live" with phase space as a fundamenal aspect of the
description and not attempt to derive it or do away with it from a purely
space-time based approach.

\section{Phase Spaces and Groups}

From \textit{classical} experiments, one learns that classical (Newtonian)
equations of motion are invariant under translations, boosts (relative
velocity transformations between inertial [Galileian] reference frames), and
rotations. Prior\ to 1887, these were invariably viewed to generate the
group of Galileian transformations on spacetime. However, since the
Michaelson-Morley\cite{M+M} experiment, and the subsequent analysis of Voigt%
\cite{Voigt}, Lorentz\cite{Lor}, Hertz's\cite{Hertz} clarification of
Maxwell's equations, the analysis of FitzGerald\cite{Fitz}, Poincar\'{e}\cite%
{Poin}, Einstein\cite{Ein}, and \linebreak Minkowski\cite{Mink}, these
spacetime translations, boosts and rotations were henceforth interpreted as
the generators of the group of Lorentz transformations on either
energy-momentum space or on spacetime. These transformations generated the
entire group (known either as the inhomogeneous Lorentz group or the Poincar%
\'{e} group) from those transformations acting on an arbitrarily small
neighborhood of any point (i.e. those transformations in an arbitrarily
small neighborhood of the identity in the group). Transformations \textit{%
infinitesimally} near the identity transformation form a vector space (the
Lie algebra of the group) on which a non-associative operation (the Lie
bracket) is defined. This was already well-known at the time to
mathematicians (Lie, Poincar\'{e}, and others). Thus,

$\bullet $\ Classical experiments reveal the relevant kinematical groups.

The lesson learned through the efforts of mathematicians over the last fifty
years is that

$\bullet $ Classical mechanics is describable mathematically on a space with
a Poisson bracket, a \textit{phase space,} or more particularly on a
symplectic manifold which possesses a closed, non-degenerate 2-form on it.
Furthermore, the relevant Galilei or Poincar\'{e} group acts on this space
in such a way as to preserve the Poisson bracket (acts "symplectically"). A
necessary consequence of this set-up is that so-called "conjugate variables"
arise naturally; these are coordinates on the phase space which realize the
canonical skew-symmetric form of the Poisson bracket.

With the experience of the Galilei and Poincar\'{e} groups, one may abstract
this formulation to the setting of the action of a Lie group on \textit{any}
phase space.

The group $G$, being a Lie group, possesses an associated Lie algebra $%
\mathfrak{g}$ that may be thought of as the collection of all left-invariant
vector fields on \textit{G}. There is a formal invertible process of e%
\textit{xponentiation }that associates an element of the group (near the
identity) to any element of the Lie algebra sufficiently near the origin
(zero). One may thus go from the Lie algebra to the Lie group, and \textit{%
vice versa}. In what follows it is essential that $\mathfrak{g}$\ is a
finite-dimensional vector space. If $\wedge $\ designates the anti-symmetric
tensor product on $\mathfrak{g}$ then one may form the skew-symmetric tensor
algebra $\mathcal{\Lambda (}\mathfrak{g)}$ over $\mathfrak{g}$ consisting of
elements of various types, namely: $\mathbb{R}$, $\mathfrak{g}$, $\mathfrak{%
g\wedge g}$, $\mathfrak{g\wedge g\wedge g}$, etc. Let their duals be denoted
by $\mathfrak{g}^{\ast }$, etc. and note that $\mathfrak{g}^{\ast }$ may be
thought of as the collection of all left-invariant 1-forms on $G$, $%
\mathfrak{g}^{\ast }\mathfrak{\wedge \mathfrak{g}^{\ast }\ }$as the
left-invariant 2-forms on $G$, and so on. One defines the coboundary
operator $\delta $ 
\[
\mathbb{R}\longrightarrow ^{\delta _{0}}\mathfrak{g}^{\ast }\longrightarrow
^{\delta _{1}}(\mathfrak{g}\wedge \mathfrak{g})^{\ast }\longrightarrow \cdot
\cdot \cdot 
\]%
as follows. Let $(A_{i})$ be a basis of $\ \mathfrak{g}$ and let ($\omega
^{i})$ be the associated dual basis of \ $\mathfrak{g}^{\ast }$\ so that $%
\omega ^{i}(A_{j})=\delta _{j}^{i}.$ The structure constants of $\mathfrak{g}
$,\ defined relative to the basis $(A_{i})$, are determined by the Lie
bracket relations: [$A_{i},A_{j}]=\dsum\limits_{k}C_{ij}^{k}A_{k}$. The $%
\mathbb{R}$ in the sequence above can be considered to be the collection of
left-invariant functions on the group $G$, which is assumed to be connected,
so that the $\mathbb{R}$ may be thought of as the left-invariant 0-forms $f$
on the group. We define%
\[
\delta _{0}f=0 
\]%
as an element of $\mathfrak{g}^{\ast }$. Now, thinking of the $\omega ^{i}$\
as left-invariant 1-forms one finds that the Maurer-Cartan equations hold: d$%
\omega ^{k}=-\frac{1}{2}\dsum\limits_{i,j}C_{ij}^{k}\omega ^{i}\wedge \omega
^{j}$. We then define 
\[
\delta _{1}\omega ^{k}=-\frac{1}{2}\dsum\limits_{i,j}C_{ij}^{k}\omega
^{i}\wedge \omega ^{j} 
\]%
recognizing that this 2-form is actually in $(\mathfrak{g}\wedge \mathfrak{g}%
)^{\ast }$. One extends this expression for $\delta _{1}$ linearly and
thereby obtains the linear map $\mathfrak{g}^{\ast }\longrightarrow ^{\delta
_{1}}(\mathfrak{g}\wedge \mathfrak{g})^{\ast }$. Making use of the
skew-derivation property for $\delta _{2}$ 
\[
\delta _{2}(\lambda \wedge \mu )\equiv (\delta _{1}\lambda )\wedge \mu
-\lambda \wedge (\delta _{1}\mu ), 
\]%
for $\lambda ,\mu \in \mathfrak{g}^{\ast }$, one defines $\delta $
inductively.

Letting 
\[
Z^{2}(\mathfrak{g})\equiv \left\{ \omega \in (\mathfrak{g\wedge g)}^{\ast
}\mid \delta _{2}(\omega )=0\right\} 
\]%
denote the space of closed, left-invariant 2-forms on $G,$ for $\omega \in
Z^{2}(\mathfrak{g})$, define

\[
h_{\omega }\equiv \{\xi \in \mathfrak{g}\mid \omega (\xi ,\cdot )=0\}. 
\]%
Then $h_{\omega }$\ is a Lie sub-algebra of\ $\mathfrak{g}$ and $h_{\omega }$%
\ determines, by exponentiation, a subgroup $H_{\omega }$\ of $G$. Supposing
that $H_{\omega }$\ is a closed subgroup of $G$, 
\[
\Gamma \equiv G/H_{\omega } 
\]%
is a manifold. That it is a symplectic manifold (of even dimension equal to
2m for some integer m) follows from the fact that the 2-form $\omega ,$ when
factored by its kernel, is the pull-back of a non-degenerate closed 2-form
on $G/H_{\omega }$. That it is a symplectic $G$\ space follows because $G$
acts on $G/H_{\omega }$ by left multiplication on left cosets: $%
gx=g(g_{1}H_{\omega })=(gg_{1})H_{\omega }$, where $x=g_{1}H_{\omega }$ for
some $g_{1}$ in $G$. Since $\Gamma \equiv G/H_{\omega }$ is a symplectic
manifold, it naturally possesses a left-invariant Liouville measure\ $\mu $\
equal to the m-th exterior power of $\omega $.

The following result (Theorem 25.1 of\cite{GSt}) captures the essence of the
need for the construction outlined above and is sufficient for our purposes,
but only in the context of \textit{single-particle} kinematics.

\begin{theorem}
Any symplectic action of a connected Lie group G on a symplectic manifold M
defines a G morphism, $\Psi :M\rightarrow Z^{2}(\mathfrak{g})$. Since the
map $\Psi $ is a G morphism, $\Psi (M)$ is a union of G orbits in $Z^{2}(%
\mathfrak{g})$. In particular, if the action of G on M is transitive, then
the image of $\Psi $ consists of a single G orbit in $Z^{2}(\mathfrak{g})$.
\end{theorem}

In the case of the inhomogeneous Lorentz group the co-adjoint orbit
construction is sufficient, whereas in the case of the inhomogeneous Galilei
group one must consider the symplectic cohomology groups \textit{H}$^{1}(%
\mathfrak{g}$) and \textit{H}$^{2}(\mathfrak{g}$) which are both
non-trivial. One may consult section 25 of \cite{GSt}.

$\bullet $ In this fashion, one obtains ALL the \textit{single-particle }%
symplectic spaces on which $G$\ acts symplectically and transitively. In
consequence one has a unified mathematical picture of kinematics in the two
fundamental cases (Galileian and Lorentzian) of relevance to one-particle
physics. Multi-particle kinematics is then described by a phase space that
is a Cartesian product of the single-particle phase spaces with symplectic
form equal to the "sum" of the symplectic forms on each of the
single-particle factors. This is the "m\'{e}thode de fusion"\cite{Sour}. In
other words, starting from the symplectic action of a group on classical
single-particle phase space, one obtains all the phase spaces (single- or
multi-particle) on which $G$ acts symplectically, in a physically meaningful
way.

$\bullet $ The coordinates on each of these spaces can be sorted into the
momentum, position, and rotation coordinates for massive particles, or the
frequency, position, and helicity coordinates in the case of the zero mass
particles. In answer to the question: "Where does one get these canonical
coordinates?" asked by David Finkelstein\cite{Finkel}, this discussion
provides at least a partial answer.

$\bullet $ It is emphasized that the same procedure will work for any
(connected) Lie group. Thus, results for the Heisenberg group, the affine
group, the de Sitter group, etc. have been obtained.

\section{Hilbert Space Associated to Phase Space}

Having chosen $\omega \in Z^{2}(\mathfrak{g})$, and obtained $\Gamma
=G/H_{\omega }$ and $\mu $, one may form $L_{\mu }^{2}(\Gamma )$, which is a
Hilbert space on which one may represent $G$\ by unitary operators $V(g)$%
\[
\lbrack V(g)\Psi ](x)\equiv \Psi (g^{-1}x) 
\]%
for $\Psi \in L_{\mu }^{2}(\Gamma )$. Note that it may be necessary in some
situations to extend the representation above by incorporating a phase
factor.

One may define an operator $A(f),$\ for all $\mu $-measurable $f$, by%
\[
\lbrack A(f)\Psi ](x)\equiv f(x)\Psi (x). 
\]%
These operators on $L_{\mu }^{2}(\Gamma )$\ have, in the case where the $f$
are characteristic functions $\chi (\Delta )$, the clear
classically-motivated interpretation of localization observables in the
phase space region $\Delta $. The collection of quantum mechanical
observables includes non-commuting operators and hence must contain
operators other than operators of the form $A(f)$. It will become evident
that $L_{\mu }^{2}(\Gamma )$\ is not a Hilbert space of fundamental
importance to the description of Quantum Mechanical models of elementary
(i.e., irreducible), single-particle systems, but, that it is reducible into
a direct sum (or integral) of such irreducible spaces.

\section{Quantum Mechanical Representation Spaces}

In the case of the inhomogeneous Galilei and Lorentz groups, the "Mackey
Machine" \cite{Mack} and the earlier Wigner classification \cite{Wig39} are
well-known to produce all continuous, irreducible, unitary Hilbert space
representations and that these are characterized by the Casimir invariants
in the universal enveloping algebra of the Lie algebra. These Casimir
elements are identifiable as the physical quantities of rest mass and spin
(or helicity in the mass-zero case). For the inhomogeneous Galilei group one
had to wait until the analysis of L\'{e}vy-Leblond \cite{Levy} to achieve a
similar picture physically characterized by mass and spin.

It is the case that the well-known irreducible representation spaces for
both the Galilei and Lorentz group are "single-particle Hilbert spaces" in
the usual language of Physics, and are Hilbert spaces of square-integrable
functions over single-particle momentum-energy spaces.

What we will see later is that the correspondence between "irreducible" and
"single-particle" is best elucidated within the Hilbert space
constructed-over-phase-space framework.

In what follows $U$\ will usually denote an irreducible unitary
representation of\ $G$\ on an irreducible representation space, usually
denoted $\mathcal{H}$.

\section{Phase Space and Quantum Representations}

The critical idea is the following.

One wishes to define a linear transformation $W^{\eta }$ from $\mathcal{H}$
to $L_{\mu }^{2}(\Gamma )$ by

\[
\lbrack W^{\eta }(\varphi )](x)\equiv \text{ }<U(\sigma (x))\eta ,\varphi > 
\]%
for $x\in \Gamma =G/H_{\omega }$, for all $g\in G$, and for all $\varphi \in 
\mathcal{H}$, where $\eta $ is a vector in $\mathcal{H}$\ and where $\sigma $%
\ is a (Borel measurable) section

\[
\sigma :G/H_{\omega }\longrightarrow G. 
\]

The reason to define such a map is that one seeks to encode the entire
content of the state vector $\varphi \in \mathcal{H}$ into a complex-valued
function on the phase space $\Gamma $ in a manner that is reversible. The
goal is to be able to reconstruct the state from the complex numbers $%
[W^{\eta }(\varphi )](x)$ which encode it.

To ensure that the image of $W^{\eta }$ actually lies in $L_{\mu
}^{2}(\Gamma )$ one must exercise some care in the choice $\eta $.
Accordingly:

\qquad (a) one selects and fixes, once and for all, a (Borel measurable)
section $\sigma :G/H_{\omega }\longrightarrow G;$

\qquad (b) one chooses a "suitable" resolution generator $\eta \in \mathcal{H%
}.$

The trick here is to decide what "suitable" means. One says that $\eta $ is 
\textit{admissible} with respect to the section $\sigma $ if%
\[
\dint\limits_{\Gamma }\mid <U(\sigma (x))\eta ,\eta >\mid ^{2}d\mu
(x)<\infty . 
\]

Assuming that $\eta $ is admissible with respect to $\sigma $, one says that 
$\eta $ is $\alpha $\textit{-admissible with respect to }$\sigma $ if in
addition to admissibility of $\eta $ one also has%
\[
U(h)\eta =\alpha (h)\eta 
\]%
for all $h$ in $H_{\omega }$, where $\alpha $ is a one-dimensional
representation of $H_{\omega }.$

If $\eta $ is $\alpha $\textit{-admissible with respect to }$\sigma $ then
we have what is needed to properly define the mapping $W^{\eta }$ from $%
\mathcal{H}$ to $L_{\mu }^{2}(\Gamma )$ and to carry out the analysis needed
to describe states $\varphi \in \mathcal{H}$ by their images $W^{\eta
}(\varphi )$ in $L_{\mu }^{2}(\Gamma )$.\cite[chapter III pages 317-234 ]{S1}

To illustrate these conditions, consider:

$\bullet $ the case of a massive, spinless, relativistic particle ($G=$
Poincar\'{e} group) in which one finds \cite{Ali} that $\eta $ must be
rotationally-invariant under $H_{\omega }=SU(2),$ and square-integrable over 
$\Gamma \equiv G/H_{\omega }\cong \mathbb{R}^{6}\cong \mathbb{R}%
_{position}^{3}\times \mathbb{R}_{momentum}^{3}$ the classical phase space
of a massive, relativistic spinless particle.

$\bullet $ the case of a massive, relativistic particle with non-zero spin ($%
G=$ Poincar\'{e} group) in which one finds \cite{BS2}\cite{BrSc} that $\eta $
must be rotationally invariant about the "spin axis" (but not necessarily
invariant under all rotations in $SU(2)$), i.e., invariant under $H_{\omega
}=$ double covering of $O(2)$ $\cong $ stabilizer in $SU(2)$ of the spin
axis, and square-integrable over $\Gamma \equiv G/H_{\omega }\cong \mathbb{R}%
_{position}^{3}\times \mathbb{R}_{momentum}^{3}\times S_{spin}^{2}$ the
classical phase space of a massive, relativistic, spinning particle.

Orthogonality relations, which play a prominent role in the representation
theory of compact groups, also appear in this approach. Assuming that the
vectors $\eta _{i}\in \mathcal{H}$, $i=1,2$ are $\alpha $-admissible, one
may prove the existence of a unique, positive, invertible operator\ $C$\
such that for all $\varphi _{i}\in \mathcal{H},$ there holds an
"orthogonality relation" of the form \cite{HeSc}%
\[
\dint\limits_{\Gamma }<\varphi _{1},U(\sigma (x))\eta _{1}><U(\sigma
(x))\eta _{2},\varphi _{2}>d\mu (x)=\text{ }<C\eta _{2},C\eta _{1}><\varphi
_{1},\varphi _{2}> 
\]%
Note, in the case of a compact group, the positive operator $C$ simplifies
to a positive \textit{constant}. In fact this orthogonality relation holds
with $C$ a positive constant on any group in which there is satisfied yet
another admissibility condition - the $\beta $\textit{-admissibility
condition} - for an $\alpha $-admissible vector $\eta \in \mathcal{H}$.\ An $%
\alpha $-admissible vector $\eta $ is said to be $\beta $\textit{-admissible 
}if, when $g$ is any commutator of group elements $\sigma (x)^{-1}\sigma
(y)^{-1}\sigma (x)\sigma (y),$ then $U(g)\eta =\beta (x,y)\eta $ for some
scalar function $\beta (x,y)$. The $\beta $\textit{-}admissible\textit{\ }%
condition holds in the case of the inhomogeneous Galilei group, but not for
the Poincar\'{e} group, suggesting that Poincar\'{e} group orthogonality
relations are not expressible with the right-hand side of the form $\frac{1}{%
d}<\eta _{2},\eta _{1}>$ $<\varphi _{1},\varphi _{2}>$ for \textit{d} a
positive constant independent of $\eta _{1},\eta _{2},$ and $\varphi
_{1},\varphi _{2};$ if \textit{d} exists, then 
\[
1/d=\parallel \eta \parallel ^{-4}\dint\limits_{\Gamma }\mid <U(\sigma
(x))\eta ,\eta >\mid ^{2}d\mu (x). 
\]

For the sake of simplicity we denote the closure of the image of $W^{\eta }$
by $W^{\eta }(\mathcal{H})\subset L_{\mu }^{2}(\Gamma ).$ Let $P^{\eta }$
denote the canonical projection%
\[
P^{\eta }:L_{\mu }^{2}(\Gamma )\longrightarrow W^{\eta }(\mathcal{H}) 
\]%
and denote by $A^{\eta }(f)$\ the mapping \cite{S1}%
\[
A^{\eta }(f)\equiv \lbrack W^{\eta }]^{-1}P^{\eta }A(f)W^{\eta }:\mathcal{H}%
\longrightarrow \mathcal{H}. 
\]

This is a plausible candidate for the quantum mechanical operator that
corresponds to the classical observable $f$. For example, for the Heisenberg
group and for $\eta =$\ the ground state wave function of the harmonic
oscillator, then $A^{\eta }(q)=Q$ = the position operator, and $A^{\eta
}(p)=P$ = the momentum operator.

One can prove \cite{S1} that $A^{\eta }(f)$\ has an operator density $%
T^{\eta }(\cdot )$:%
\begin{eqnarray*}
A^{\eta }(f) &=&\dint\limits_{\Gamma }f(x)T^{\eta }(x)d\mu (x), \\
T^{\eta }(x) &\equiv &\mid U(\sigma (x))\eta ><U(\sigma (x))\eta \mid ,
\end{eqnarray*}%
and that 
\[
A^{\eta }(1)=1. 
\]

With this set-up one can make a number of remarks:

1) Let $\rho $\ denote any quantum density operator; i.e., $\rho $ is
non-negative and has trace one. Then one may write $\rho =\dsum \rho
_{i}P_{\psi _{i}},$ the $\psi _{i}$\ forming an orthonormal set and $P_{\psi
_{i}}$\ denoting the corresponding projection. Now, using the interpretation
of $\mid <U(\sigma (x))\eta ,\psi _{i}>\mid ^{2}$\ as the transition
probability from\ $\psi _{i}$\ to $U(\sigma (x))\eta $, one has the quantum
expectation value given by%
\[
Tr(\rho A^{\eta }(f))=\dsum\limits_{i}\rho _{i}\dint\limits_{\Gamma
}f(x)\mid <U(\sigma (x))\eta ,\psi _{i}>\mid ^{2}d\mu (x); 
\]%
i.e., the sum over the transition probabilities. \cite{S1}

For example, when using a "screen" to detect a particle in a vector state
given by $\psi $, one idealizes the detector (the screen) as a
multi-particle quantum system consisting of identical sub-detectors. In a
fixed laboratory frame of reference a sub-detector is represented by a state
vector $\eta $ whose phase space counterpart $W^{\eta }\eta $ is peaked
about a reference phase space point which may be referred to as "the
origin". For a fixed space-time reference frame, one may "position" a
detector at all "points" of space-time (space-time events) exactly as
Einstein located rods and clocks. Of course, one must now position mass
spectrometers (devices that measure rest-mass in their own rest frames) and
Stern-Gerlach devices at all space-time events in addition to rods and
clocks. As Einstein imagined that the rods and clocks were also equipped (at
all space-time coordinate events) in all inertially-related space-time
reference frames, so must we imagine that our inertially-related space-time
reference frames carry identical mass spectrometers and Stern-Gerlach
devices in addition to rods and clocks (boosted relative to the rest
"laboratory" frame). So, instead of rods and clocks situated at each
space-time event and at rest in inertially-related (uniformly moving) rest
frames, we must add\ to that \textit{imagery} a more elaborate set of
apparati. For a fixed value of momentum $p$ there are infinitely many pairs $%
(m,u)$ such that $p=mu$; of course the momentum does not alone characterize
the uniform relative velocity (boost) represented by $p$ - one requires also
the rest-mass $m$. The totality of all such "placements" of detectors
constitutes the phase-space distribution of detectors - the classical phase
space frame analogous to the classical space-time (Lorenz) frame (of rods
and clocks). Thus the complete detector is composed of sub-detectors each
located at different "positions" (points of $\Gamma $). The sub-detector
located at "position" $x\in \Gamma $, obtained from $\eta $ by a kinematical
placement procedure (with the same intent as Einstein's placement of
identical rods and clocks at all points of spacetime), is $U(\sigma (x))\eta 
$. Since the probability that\ $\psi $\ is captured in the state given by $%
U(\sigma (x))\eta $ is $\mid <U(\sigma (x))\eta ,\psi >\mid ^{2}$, the
formula for the expectation is justified. \textit{One cannot improve upon
this procedure when measuring, by quantum mechanical means, the distribution
of the particle. }

2) Since $T^{\eta }(x)$ $\geq 0$ and $A^{\eta }(1)=1$, then $\rho
_{class}(x)\equiv Tr(\rho T^{\eta }(x))$\ is a classical (Kolmogorov)
probability function \cite{S1}. Consequently,%
\begin{eqnarray*}
\text{quantum expectation } &=&Tr(\rho A^{\eta }(f)) \\
&=&\dint\limits_{\Gamma }f(x)Tr(\rho T^{\eta }(x))d\mu (x) \\
&=&\dint\limits_{\Gamma }f(x)\rho _{class}(x)d\mu (x) \\
&=&\text{ classical expectation.}
\end{eqnarray*}

3) Since the operators $A^{\eta }(f)$ enjoy the feature of the same
expectation as the "classical" obseverables $f$, one might ask whether these
operators are sufficient to distinguish states of the quantum system.

\begin{definition}
\cite{Prug0}A set of bounded self-adjoint operators $\{A_{\beta }\mid \beta
\in I,$ $I$ some index set$\}$ is informationally complete iff for all
states $\rho ,\rho ^{\prime }$ such that $Tr(\rho A_{\beta })=Tr(\rho
^{\prime }A_{\beta })$ for all $\beta \in I$\ then $\rho =\rho ^{\prime }$.
\end{definition}

Example\cite{Prug0}: In spinless quantum mechanics, the set of all spectral
projections for position is not informationally complete. Neither is the set
of all spectral projections for momentum, nor even the union of them.

The $\{A^{\eta }(f)\mid f$ is measurable$\}$\ (or, equivalently $\{T^{\eta
}(x)\mid x\in \Gamma \}$) is known to be informationally complete in a
number of cases:

\qquad a) spin-zero massive representations of the Poincar\'{e} group\cite%
{Ali}

\qquad b) mass-zero, arbitrary helicity representations\cite{BS} of the
Poincar\'{e} group

\qquad c) the affine group \cite{HeSc}

\qquad d) the Heisenberg group \cite{S1}

\qquad e) massive representations \cite{AlPr}\cite{S1} of the inhomogeneous
Galilei group

\qquad f) massive, non-zero spin representations\cite{BrSc} of the Poincar%
\'{e} group are being investigated

4) If $\{A_{\beta }\mid \beta \in I\}$\ is informationally complete then any
bounded operator on $\mathcal{H}$ may be written as (a closure of) integrals
over the set $I$.\cite{Bu}

5) When we specialize $A^{\eta }(f)$\ to $f=\chi (\Delta )$\ , $\chi (\Delta
)$\ the characteristic function for the Borel set $\Delta $, then%
\begin{eqnarray*}
\chi (\Delta ) &=&\text{classical localization in }\Delta \subset \Gamma 
\text{,} \\
A(\chi (\Delta )) &=&\text{operator on }L^{2}(\Gamma )\text{ localizing in }%
\Delta \subset \Gamma \text{,} \\
A^{\eta }(\chi (\Delta )) &=&\text{operator on }\mathcal{H}\text{ localizing
in }\Delta \subset \Gamma \text{.}
\end{eqnarray*}

These $A^{\eta }(\chi (\Delta ))$\ have several properties \cite{S1}:

$\qquad $a) If $\Delta $ is a compact subset of $\Gamma $ then $A^{\eta
}(\chi (\Delta ))$\ is a compact operator with spectrum in [0,1],

$\qquad $b) For all $\Delta \subset \Gamma ,$ $\parallel A^{\eta }(\chi
(\Delta ))\parallel $ $\leq \mu (\Delta ).$

Some consequences of this set-up appear in subsequent sections.

\section{Comment on a paper by Carmeli, Cassinelli, Toigo, and Vacchini}

The paper in question is: \textit{A complete characterization of phase space
measurements. }\cite{Car}

The following remark holds also for the case of non-zero spin for either
Galileian or Lorentzian massive one-particle situations, but is simplified
to the case treated in the paper by Carmeli \textit{et al}, namely,
spin-zero.

In the present formalism, by making use of an isometry (the Wigner
transform) from the Hilbert space of square-integrable functions on
three-dimensional space (the irreducible representation space of the
canonical commutation relations [of the Weyl-Heisenberg group]) to a
subspace of the Hilbert space of square-integrable functions on physical
six-dimensional phase space, one may determine the irreducible unitary
subrepresentations of the inhomogeneous Galilei and Lorentz groups arising
from a \textit{resolution generator} via the Wigner transform.\cite{Prug2}%
\cite{Prug3}. See also \cite{AliPru}\cite{Br}. In this way, the invariance
of the density matrix under rotations (as posited by Carmeli, \textit{et al}%
) results from the invariance of the resolution generator under rotations.
The advantage of the resolution generator view is that the localization
operators on phase space arise naturally within the unified theory in which
the quantum mechanical Hilbert space is constructed directly from the
classical phase space. Moreover, the same construction applies in the
relativistic setting, and furthermore, in the non-zero-spin situations.\cite%
{Ali}

Carmeli, \textit{et al }treat spin-zero, massive, one-particle,
non-relativistic quantum mechanics and obtain a characterization of phase
space measurements. The authors state in section 3: "In the present section,
we characterize all the phase space measurements of a non-relativistic
particle of mass \textit{m}. For the sake of simplicity we restrict to the
spinless case, the extension to the general case being straightforward." To
the contrary, it is our experience that the non-zero spin case is not as
straightforward as is often claimed IF, as outlined above, one is expected
to introduce the spin through phase space and group theoretic considerations
rather than by \textit{ad hoc} constructions. They have it backwards from
the present point of view in the sense that the phase space is determined by
the kinematic group. They do not take into account the fact that the spin
and the\ angular momentum are intertwined. Their proof of their principal
result may not be valid in the relativistic situation where, as was
mentioned in Section V, in the Poincar\'{e} case one should not expect an
orthogonality relation with the operator $C$ a constant equal to 1/$d.$

\section{Uncertainty Relations \& Channel Capacity Theorem}

Consider the measuring instrument (represented by $\eta $) to be fixed.
Since $A^{\eta }(\chi (\Delta ))$ is a compact operator for compact $\Delta $
\cite{S1}, let it have eigenvalues $\lambda _{i}$\ with corresponding
eigenvectors $\psi _{i}:$%
\[
A^{\eta }(\chi (\Delta ))\psi _{i}=\lambda _{i}\psi _{i},\text{ }0\leq
\lambda _{i}\leq 1. 
\]%
One says that $\psi _{i}$ is \textit{localized in }$\Delta $ if $\lambda
_{i} $ is \textit{close} to 1 (say $\lambda _{i}>1-\epsilon $). When
localized (in $\Delta $) by $A^{\eta }(\chi (\Delta ))$, $\psi _{i}$\ will
be attenuated by the factor $\lambda _{i}$. One has from the above that 
\begin{eqnarray*}
\lambda _{i} &=&Tr[A^{\eta }(\chi (\Delta ))\mid \psi _{i}><\psi _{i}\mid ]
\\
&=&\dint\limits_{\Delta }Tr[T^{\eta }(x)\mid \psi _{i}><\psi _{i}\mid ]d\mu
(x)\leq \mu (\Delta )
\end{eqnarray*}%
and, in fact, more strongly that

\[
\dsum\limits_{i}\lambda _{i}=Tr[A^{\eta }(\chi (\Delta ))]\leq \mu (\Delta ) 
\]%
which is a sharper upper bound on the $\lambda _{i}$ when $\mu (\Delta )$ is
small, in particular if $\mu (\Delta )\leq 1,$ in units of $\hslash ,$ when
the phase space is two-dimensional.

This exemplifies the following version of "the uncertainty relation": it is
impossible to localize a physical quantum system in an arbitrarily small
volume in phase space\cite{Feyn}. The following result is useful to
establish this uncertainty relation. If $\Delta $ has a smooth boundary, one
can prove \cite{S1} that the $\lambda _{i},$ are clustered near 1 and 0 with
almost no $\lambda _{i}$ between $\epsilon $ and $1-\epsilon ,$ for some $%
\epsilon >0$. (For example, take $\epsilon =\frac{1}{n}$, where $n$ is an
integer bigger than 3.) If there were $N$ of the $\lambda _{i}$ clustered
near 1 then $N\approx \dsum\limits_{\lambda _{i}\text{ near 1}}\lambda
_{i}\leq \mu (\Delta )$ which, when $\mu (\Delta )$ is small, requires $%
N\leq 1$.

There are a number of examples from the world of classical mechanics whose
analysis is improved by treating it as a quantum mechanical system. Our
first example is the channel capacity theorem: "In a time interval from $-T$%
\ to $T$\ and in a band width of size $\Omega $, the total number of
channels that can pass through the device is $2\Omega T$".\ This theorem was
originally argued to be true by "modifying" the signal and its Fourier
transform. But we know mathematically, that a non-zero signal and its
Fourier transform may not both have compact supports. Now, if we take the
time-frequency space as a phase space and treat the number of channels as
the number of orthogonal wave functions $\psi _{i}$\ that can pass though
(read as "when localized in") the device without severe attenuation, we can
obtain the "$2\Omega T$" result from the analysis above.

There are many other subjects that can be profitably analyzed with this
phase space formalism. It may seem strange to consider some of them as
quantum mechanical systems, but that has been done. To list a few: 1)
neutron interferometry, 2) single slit experiments, 3) Stern-Gerlach
devices, 4) CT scans, 5) N.M.R., 6) M.R.I., 7) holography, 8) bat
echo-location, 9) the olfactory system of dogs, 10) neural networks in the
brain, 11) geologic exploration, 12) clearing mine fields (which is being
investigated by someone at this conference), 13) radar, etc. Many of these
and others have been investigated and results may be found in \cite{Busch}%
\cite{S1}.

To make \textit{obvious} the point that the phase space perspective is
necessary, take the system by which one \textit{sees}. One's brain creates a
display of both the image in 3-space and in color. The phase space of the
photon of either positive or negative helicity is topologically homeomorphic
to $\mathbb{R}^{3}\times \mathbb{R}^{+}\times S^{2}$\ where the $\mathbb{R}%
^{3}$-factor is the position space, the $\mathbb{R}^{+}$-factor is the
frequency space, and the $S^{2}$-factor\ is the space of rotations in the
momentum space. See \cite{BS}. Thus, you may place an instrument at any
point in configuration space, turn it so it points in any direction, and
then measure the frequency (or wave-length) and helicity. When one looks in
one direction (possibly aided by polarized glasses), the brain makes
measurements in phase space!

\section{Effect Algebras}

To bring the discussion closer to other of the topics of this conference, we
begin with three definitions.

\begin{definition}
An operator $A$ in any Hilbert space is an \textbf{effect }if $A$\ is
self-adjoint, non-negative ("positive"), and bounded above by 1.
\end{definition}

\begin{definition}
An \textbf{effect algebra} $E$ is a set containing 0 and 1, with a partial
binary operation $\oplus $ on $E$ satisfying i) if $a,b$ and $a\oplus b\in E$%
, then $b\oplus a\in E$ and $a\oplus b=b\oplus a;$ \ ii) if $a,b,c,b\oplus
c,a\oplus (b\oplus c)\in E,$ then $a\oplus b,(a\oplus b)\oplus c\in E$ and $%
a\oplus (b\oplus c)=(a\oplus b)\oplus c;$\ iii) $\forall $ $a\in E,\exists !$
$a^{\prime }\in E$ such that $a\oplus a^{\prime }=1;$ iv) if $a\oplus 1\in
E, $ then $a=0.$
\end{definition}

The set of all effects in a Hilbert space is \textit{an} effect algebra. As
one will see, in a Hilbert space it is not the only one.

If\ $A$\ is an effect on $\mathcal{H}$ and\ $\rho $\ is any density operator
on $\mathcal{H}$, then $0\leq Tr(A\rho )\leq 1$. Thus one may view $Tr(A\rho
)$\ as the expected value of $A$\ in state $\rho $.

\begin{definition}
A positive operator-valued measure (POVM) is a mapping $A$ from any $\sigma $%
-algebra $\Sigma $\ on any set\ $\Gamma $\ to the non-negative ("positive")
self-adjoint operators on $\mathcal{H}$ such that: i) $A(\Gamma )=1$, (and $%
A(\phi )=0$), ii) for every countable collection $\{\Delta _{i}\}$\ of
disjoint measurable sets ($\Delta _{i}\in \Sigma $), $A(\cup _{i}\Delta
_{i})=\Sigma _{i}A(\Delta _{i})$\ (in the topology of weak operator
convergence). A projection-valued measure (PVM) is a POVM in which all $%
A(\Delta _{i})$\ are projections.
\end{definition}

If, in the formalism of Quantum Mechanics on Phase Space, one defines%
\begin{eqnarray*}
E^{\eta } &\equiv &\{A^{\eta }(f)\mid 0\leq f\leq 1,\text{ }f\text{ Borel
measurable}\}\text{,} \\
F^{\eta } &\equiv &\{A^{\eta }(\chi _{\Delta })\mid \Delta \text{ Borel
measurable on }\Gamma \},
\end{eqnarray*}%
with $\oplus $\ defined by $A^{\eta }(f_{1})\oplus A^{\eta }(f_{2})\equiv
A^{\eta }(f_{1}+f_{2})$ when $f_{1}+f_{2}\leq 1,$ $[A^{\eta }(f)]^{\prime
}\equiv 1-A^{\eta }(f),$ etc.,\cite{S2}.\ then one obtains the following

\begin{theorem}
$F^{\eta }\subset E^{\eta };$ $E^{\eta },F^{\eta }$ are effect algebras; $%
F^{\eta }$\ generates a POVM which is informationally complete on $\mathcal{H%
}$ for suitable $\eta $.
\end{theorem}

Remarkably, we have

1) There is no projection in $E^{\eta }$\ other than $0$\ and $1$. Thus, one
does not obtain a PVM from either $E^{\eta }\ $or $F^{\eta }$.

2) $E^{\eta }$\ is not only an effect algebra, it is also an interpolation
algebra, a Riesz decomposition algebra, a\ lattice ordered effect algebra, a
distributive algebra, an M.V. algebra, and a Heyting algebra. It is not a
Boolean algebra. \cite{S3}

3) $E^{\eta }$\ is not an orthoalgebra. The property $a\wedge a^{\prime }=0$%
\ is equivalent to all the $A^{\eta }(f)$'s being projections, which is
ruled out by 1). Included in this are all "finite quantum logics." To
approximate these projections, one would need to look at an informationally
complete set of the $A^{\eta }(f)$'s for the $f$'s being just measurable
real-valued functions.

4) The $\eta $\ involved is a wave function for the measuring instrument. It
is an essential ingredient to achieve a true quantum measurement.

5) According to philosophers and logicians, the logic by which quantum
computers should be designed is an M.V. algebra that is also a Heyting
algebra. The remarks above justify this assertion.

6) Taking the $A^{\eta }(f)$s as the only realistically allowed operators
in\ the theory of quantum computers, then \textit{the theory} \textit{based
on projections} is only an approximation to reality. (Here "approximation"
is based on the fact that the $A^{\eta }(f)$'s are informationally
complete.) One must have, at a minimum, a POVM that is not a PVM.
Furthermore, that POVM must reflect phase space variables in some sense.
When one makes a finite-dimensional approximation of the Hilbert space to
carry out numerical computations, one must make an appropriate choice of a
POVM. The same can be said of any numerical computations in quantum theory.

7) Given that the $A^{\eta }(f)$s are the only realistically allowed
operators, one may re-analyze the axioms of "quantum computation" and their
consequences under which the "results" of quantum computation are derived.
For example, whether Shor's algorithm for factoring large numbers is indeed
implementable in any approximate sense should be investigated.

\section{Conclusion}

Any kinematical group may be analyzed in the fashion of Quantum Mechanics on
Phase Space. Quantum Mechanical measurement usually leads to a POVM that is
not a PVM, a direct consequence of an inherent spread of the wave function
of the particles being measured. The effects of decoherence and measurement
inaccuracy are in addition to this inherent imprecision. It is our view that
the methods of \textit{Quantum Mechanics on Phase Space} must be taken into
account in order to express predictions and to analyze experiments in
quantum theory; in particular, in order to decide whether or not quantum
computers are physically realizable within this framework.

\end{document}